\pdfoutput=1
\documentclass[11pt]{article}

\usepackage[dvipsnames]{xcolor}
\usepackage{EMNLP2022}

\usepackage{times}
\usepackage{latexsym}

\usepackage[T1]{fontenc}
\usepackage{microtype}

\usepackage{inconsolata}

\usepackage[utf8]{inputenc}
\usepackage{pgfplots}
\usepgfplotslibrary{groupplots}
\pgfplotsset{compat=1.5}

\usepackage{graphicx} %插入图片的宏包
\usepackage{float} %设置图片浮动位置的宏包
\usepackage{subfigure} %插入多图时用子图显示的宏包
\usepackage{amsmath}

\title{VLSNR:Vision-Linguistics Coordination Time Sequence-aware News Recommendation}

\author{Songhao Han\footnotemark[1]\\ Beihang University \\hshjerry@buaa.edu.cn\And
         Wei Huang\footnotemark[1]\\ Beihang University \\aaron.weihuang@gmail.com\And
         Xiaotian Luan\footnotemark[1]\\ Peking University \\1900013098@pku.edu.cn
         }

\begin{document}
\maketitle
\begin{abstract}
\footnotetext[0]{*Three authors contributed equally to this research.}
News representation and user-oriented modeling are both essential for news recommendation. Most existing methods are based on textual information but ignore the visual information and users' dynamic interests. However, compared to textual only content, multimodal semantics is beneficial for enhancing the comprehension of users' temporal and long-lasting interests. In our work, we propose a vision-linguistics coordinate time sequence news recommendation. Firstly, a pretrained multimodal encoder is applied to embed images and texts into the same feature space. 
Then the self-attention network is used to learn the chronological sequence.
Additionally, an attentional GRU network is proposed to model user preference in terms of time adequately. 
Finally, the click history and user representation are embedded to calculate the ranking scores for candidate news.
Furthermore, we also construct a large scale multimodal news recommendation dataset V-MIND. Experimental results show that our model outperforms baselines and achieves SOTA on our independently constructed dataset.\footnote{https://github.com/Aaronhuang-778/V-MIND}

% News representation and user-oriented modelling are both essential for news recommendation.  A majority of the existing methods are based on textual information, comprising news title, topic and body. Meanwhile, users' dynamic interests are paid little attention to. In fact, propensity of users to click on news may be attributed to visual and literal information. Furthermore, in comparison to single-modal content, cross-modal semantic is beneficial to enhance the comprehension of users' temporal and long-lasting interest. In our work, we propose a vision-linguistics coordinate time sequence news recommendation. To begin with, pretrained multimodal encoders are leveraged to embed images and texts into the same feature space because of their outstanding performance in visiolinguistic learning. The learning of a chronological sequence are then empowered via the self attention network. Additionally, an attentional GRU network are proposed to adequately model user preference in terms of time. Finally, we encode candidate news via a cross-modal encoder as well to sort candidate news according to the relevance to both history news and user representation. Extensive comparisons with the SOTA news recommendations on our independently constructed dataset proves effective improvement on our news recommendation.\footnote{Our code and dataset will be released.}
\end{abstract}

\section{Introduction}
The enormous amount of news available everyday makes it impossible for readers to find what they are interested in immediately \cite{okura2017embedding}. Currently, news platforms like Google News\footnote{https://news.google.com}
\begin{figure}[htb]
\centering
\includegraphics[width=\linewidth]{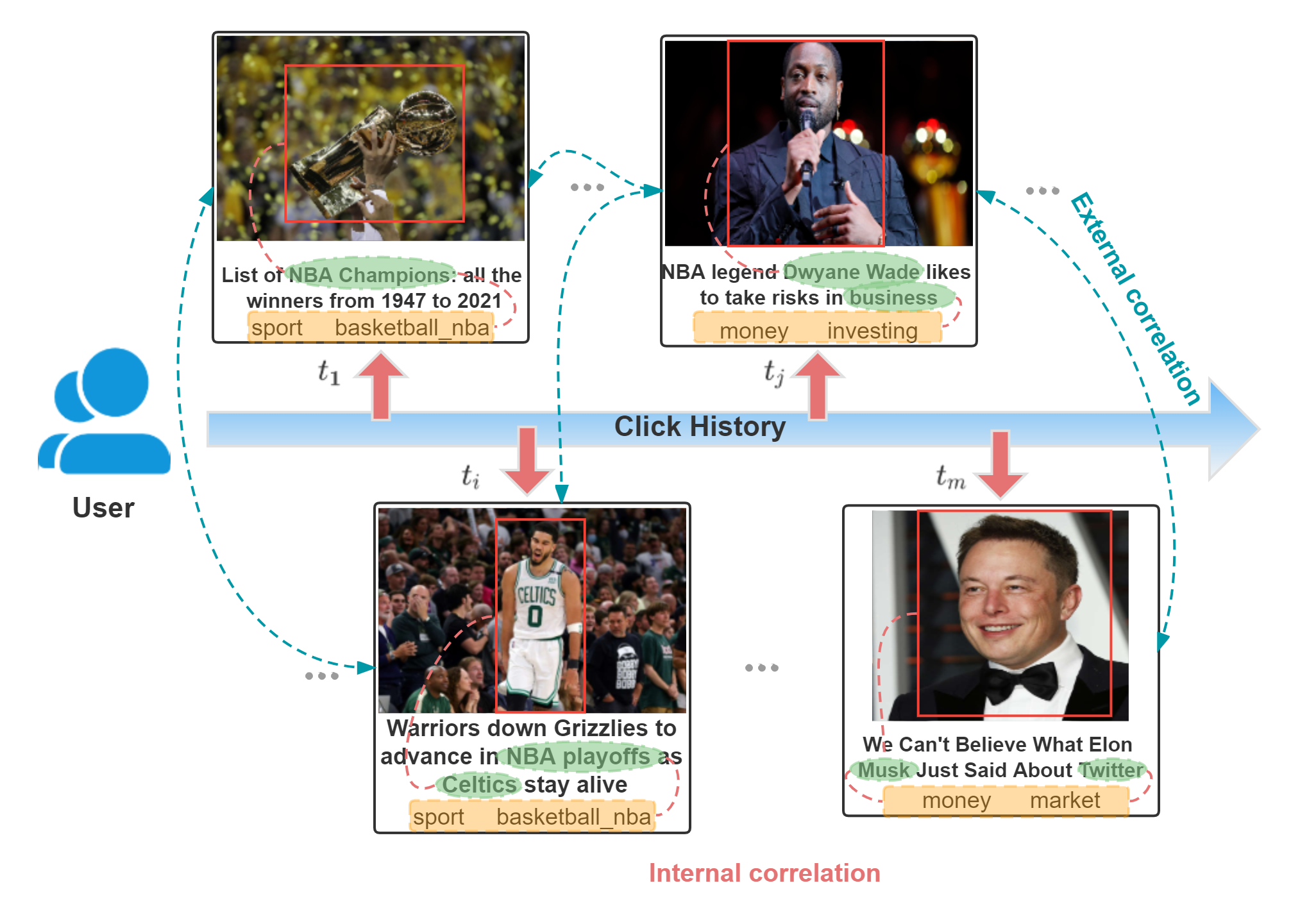} 
\caption{An illustrative example of users' dynamic interests in news with multimodal information. The blue bidirectional arrows represent external correlation. The red dotted lines represent internal correlation.}
\label{fig:1}  
\end{figure} 
and Microsoft News\footnote{https://www.msn.com/news/}
has constructed recommendation system to improve user experience.  Hence, the central target of news recommendation is to eliminate information overload and provide users with content of interest based on their habits \cite{lavie2010user}. Initially, recommendation researches concentrated on literature \cite{an2019neural,lian2018towards,wang2018dkn,wu2019npa,wu2019neural,wu2019neural2,wu2021fastformer}. For instance, \citet{okura2017embedding} proposed gate recurrent unit network to capture users' interests over time. However, it is difficult for GRU to capture contextual content information. \citet{wang2018dkn} leveraged knowledge graph with convolutional neural network to learn the interaction between history news and candidate news. The above methods are all based on text information. However, the lack of visual information may lead to inaccurate representation of users' interests. 

Recently, there are few approaches of multimodal recommendation taking texts and images into consideration \cite{wu2021mm,xun2021we}, which is closer to people's reading routine in the real world. For example, \citet{wu2021mm} propose a multimodal encoder to learn news representation from title and image, and learn users' representation using a cross-modal candidate-aware network. However, this approach ignores the significance of time sequence for user modelling. 

Our method is motivated by the fact that user interests are not only dynamic (sometimes the changes may be subtle) but also reflected in the multimodal information of the news. User's clicks are influenced by his long-lasting interests \cite{li2014modeling} and current hot spots. For example, if a user is an NBA fan, as illustrated in Figure \ref{fig:1}, this individual tends to browse much news about NBA games and players, like "NBA champions", "Celtics" and "NBA playoff" etc. elements. This kind of click tendency mainly depends on the user's long-term interest \cite{wu2019neural}. Also, the user may click some news triggered by recent hot spots, part of which are related to previous concerns. For example, in Figure \ref{fig:1}, the browsing of the news on "Wade likes to take risks in business" and title causes the user to read "Musk" and "Twitter", even if the one had never clicked on Twitter-related news before. It is noteworthy that information from different modalities is complementary or reinforcing to each other. Specifically, within a news article, the trophy in the image of the $t_1$ corresponds to the "champion" in the title, and both the topic and subtopic enhance this critical information. Among different news, the image of $t_j$ (an example of current hot-spots) also correlates with the reader's long-term interest, whereas the content is highly relevant to that of $t_m$, which can be proved by their same topic.

In this paper, we propose a Vision-Linguistics coordination time Sequence-aware News Recommendation named VLSNR, which takes advantage of fusion modules to process cross-modal information in time series. To be more precise, we establish user models via above-mentioned time-awareness network evaluated by the correlation between history clicks and candidates, which help to comprehend users' variable interest. In our approach, we transmit the image and title into CLIP encoder \cite{radford2021learning} to learn the representation of news. This enables the semantics of texts and images to be well mapped in the same feature space. Then, we construct a series of attention layers, which helps to detect deeper interaction between images and texts. Additionally, we propose an attentional GRU network\cite{okura2017embedding} to learn users' chronological interests.

%这部分删掉一块，跟contribution重复的
Noteworthy, we construct a complete dataset named V-MIND based on MIND, a benchmark data for text-domain news recommendation. Our experiments on this completely new dataset have demonstrated our approach can effectively improve the performance and our modelling can be proved to be more consistent with the behaviour of real users. Our contributions of this work are displayed as follows:
\begin{itemize}
\item[$\bullet$] We propose the VLSNR framework, which integrates visual and textual information about news in time series to learn click-preferences trends. VLSNR is able to fully characterize the overall news features and simulate users' behavior better.
\item[$\bullet$] We contribute visual information to the MIND dataset named V-MIND, which is the largest scale. It helps facilitate the future research of news recommendations in multimodal domain and improve the learning results of VLSNR. 
\item[$\bullet$] We have conducted extended experiments on the V-MIND dataset. The result shows that our method can effectively detect the interaction of visiolinguistic information and our model outperforms other news recommendation approaches.
\end{itemize}

\section{Related Work}
\begin{figure*}[htb]
\centering
\includegraphics[width=.95\linewidth]{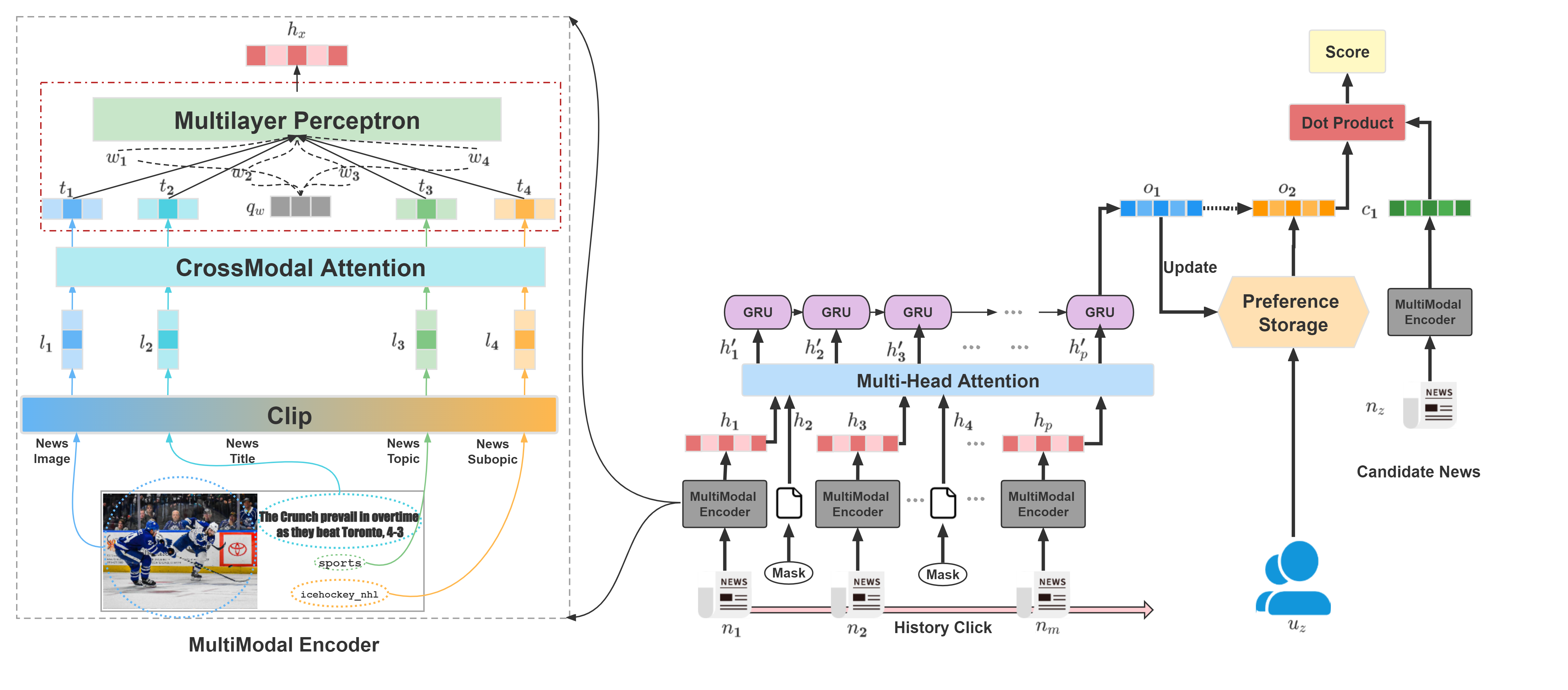} 
\caption{The model architecture of VLSNR} %图片标题
\label{fig:3}
\end{figure*}
The rapid growth of information has made news more diverse, and people are willing to spend more time browsing news. Personalized news recommendation \cite{zheng2018drn} has become an important task in the field of natural language processing and data mining and has benefited from great success in semantic analysis \cite{wang2018dkn, bahdanau2014neural,li2020unsupervised,ren2015faster} and user preference estimation \cite{okura2017embedding}. 

The processing of textual information extraction such as news headlines and topics has evolved through CNN \cite{wang2020fine}, RNN \cite{okura2017embedding} , GNN \cite{hu2020graph}and Attention mechanisms \cite{wu2019neural,zhu2019dan}. Wu et al. \cite{wu2019neural2} used the multi-head attention mechanism to expand the textual semantic information of news to uncover the relationship between the records of users' historical clicks. An et al. \cite{an2019neural} proposed two models that can characterize long-term short-term interest by combining news text features and the temporal cycle of user behavior. This is more in line with the user's behavioral habits and makes the overall effectiveness of the recommendation model improved. However, these approaches target recommendations entirely on the processing of textual features, ignoring other behavioral drivers when users click on news. News cover images have a greater visual share in news information, and the vision information of the images has an impact on user behavior.

As an efficient method, multimodal technology \cite{anderson2018bottom,zhang2020devlbert} can fuse different kinds of information such as text, images and speech to extract higher-level semantic features of news. Multimodal fusion using mainly cover information and text information in news recommendation. Wu et al.\cite{wu2021mm} used a ViLBERT \cite{lu2019vilbert} pre-training model to semantically match images and news headlines, then embedding them separately and feeding both information into the attention layer of the crossover model for prediction. Xun et al. \cite{xun2021we} which extracted the semantics of news images and news headlines with local features and used an additive attention \cite{wu2021fastformer} mechanism for multimodal fusion. Since there is no news dataset supporting multimodality, Xun's team retrieved its original news URLs based on MIND News \cite{wu2020mind} and added visual information to 54,421 news items out of 130,379 news items. These multimodal models do not combine the characteristics of users' long-period and short-period preferences for image and text information, and use a strong asymmetry in the dataset, with only 41.7\% of the image data despite the supplementation, which has an impact on the overall performance and robustness of the multimodal integrated model. In this paper we propose a recommendation architecture for long and short period interest preferences based on the fusion of news text information and image information features, and extract multimodal features using the more powerful Clip \cite{radford2021learning} model. We also crawl relevant news image covers based on news text information in Google so that the image coverage of the MIND dataset reaches 100\%.
\section{Our Method}
In this section, we introduce our \textbf{V}ision-\textbf{L}ingui-stics coordination time \textbf{S}equence-aware \textbf{N}ews \textbf{R}ecommendation approach, whose overall framework is shown as Figure \ref{fig:3}. Our structure is made up of three components, i.e., a multimodal news encoder to learn and integrate visiolinguistic representations of news, a time-series network to capture temporal characteristics and a module for users' representations. Then, we detailed present each unit.
\subsection{Multimodal News Encoder}
It is a common perception that people focus on the headline of a news item before clicking on it. However both news cover and text information affect the possibility of users to click on the news \cite{pounds2012multimodal}. Therefore, we designed a cross-modal encoder in our VLSNR structure to extract user's comprehensive information understanding of the news. Based on the information presentation of news items, we use news covers, headlines, topics and subtopics.

Visual and language-based multimodal models can effectively map low-dimensional text and images to high-dimensional features. Then it capture the semantics of text and images in a high-dimensional metric feature space. Therefore, we use Clip \cite{radford2021learning} pretrained model to capture the potential interaction features of news text and images.  For the input images, we chose ViT-B/32 \cite{dosovitskiy2020image} for processing, which encodes the images using a multi-headed attention structure similar to Bert \cite{devlin2018bert}, and Clip takes the input text and uses several Transformers \cite{vaswani2017attention} to understand the sequential semantic information of the words. Then the above image and text features are linearly projected into the multi-modal embedding space. The output is defined as $\textbf{L} = [\textbf{l}_1,\textbf{l}_2,\textbf{l}_3,\textbf{l}_4]$. These four long vectors are fed into a layer of crossmodal attention to regulate the focus of different information. The focus of the $t_{i}$ vector learned by \textbf n attention heads is computed as:
\begin{equation}
    \alpha_{i}^k={softmax}({\textbf{l}_i^T\textbf{Q}_k\textbf{L}})
\end{equation}
\begin{equation}
    \textbf{t}_i = concat(L({{\alpha}_i^1})^T;L({{\alpha}_i^2})^T ...;L({{\alpha}_i^n})^T) \textbf{V}
\end{equation}
where $\textbf Q_k$ and $\textbf{V}$ are the net parameters in each self-attention head, and ${\alpha}_i^k[j]$ is the $k^{th}$ head attention weight of $\textbf{l}_i$ on $\textbf{l}_j$. The output of $\textbf t_i$ is the result  of $\textbf{l}_i$'s crossmodal attention allocation.
$\textbf{T} = [\textbf{t}_1,\textbf{t}_2,\textbf{t}_3,\textbf{t}_4]$ has the same feature space properties, but is composed with image and text information. To perform multimodal fusion of these four vectors, we use a additive attention layer to compute their normalized weight:
\begin{equation}
    \alpha' = \textbf{q}_a^T tanh(W_a\times\textbf{T}+\textbf{b}_a)
\end{equation}
\begin{equation}
    \alpha' = {softmax}{(\alpha')}
\end{equation}
where $\textbf{q}_a$ is the query vector of additive attention, and $W_a$ and $\textbf{b}_a$ are the net parameters.
The final four information will be fused by the weights output of softmax, computed as:
\begin{equation}
    \textbf{x} = \alpha'\textbf{T}
\end{equation} 
where $\textbf{x}$ is the input of the next layer. To compress the vector space to a suitable size, we use a multilayer perceptron as the last process for feature learning,  formulted as:
\begin{equation}
    \textbf{h}_x = \textbf{g}(\textbf{W}_{i} \times\textbf{x}_i+\textbf{b}_i)
\end{equation} 
where $\textbf W_{i}$ and $\textbf b_{i}$ are the weights parameters, $\textbf x_i$ is the input of each layer in multilayer perceptron and $\textbf g$ is the activation function. The final fusion vector is expressed as $\textbf h_x $.
\subsection{Time Sequence-aware Network}
The sequential network is totally used to precisely model users' interest in chronological order according to their previous clicked histories. Take a user for instance, with the Batman movie becoming a worldwide hit, even if he is not a big fan of superhero-themed films, he may probably be attracted by relevant article, which can be attributed to his friends' experience. It is professional for GRUs\cite{cho2014learning} to capture \textbf{temporal} correlation among his browsing\cite{okura2017embedding}. To better simulate the actual situation, We have randomly added masked news to represent the noise during the browsing process. In the real world, a hockey fan might tend to follow hockey games or news about his favourite players for a \textbf{long period of time}. It is worth mentioning that we have applied attention modular to enhance global information to eliminate the negative influence of forgetfulness of earlier behaviours in a long sequence. Up to this point, both global and short-time information are taken into account. We use a history sequence $\textbf{H} =[\textbf{h}_1,\textbf{h}_2,...,\textbf{h}_p]$ to donate the representations of news. In order to detect the interactions among history news, we leverage self-attention network layer\cite{vaswani2017attention}. The outputs matrix $\textbf{H}'=[\textbf{h}'_1,\textbf{h}'_2,...,\textbf{h}'_p]$ are computed as follows:
\begin{equation}
\begin{split}
    &\beta_{i}^k={softmax}({\textbf{h}_i^T\textbf{Q}'_k\textbf{H}})
\end{split}
\end{equation}
\begin{equation}
\begin{split}
    &\textbf h'_i = concat(H({{\beta}_i^1})^T;H({{\beta}_i^2})^T ...; H({{\beta}_i^n})^T) \textbf V
\end{split}
\end{equation}
where $\textbf{Q}'_k$ is an attention query matrix, \textbf{V} is a parameter matrix.After that, the outputs matrix is sliced into vectors sequence $[\textbf{h}'_1,\textbf{h}'_2,...,\textbf{h}'_p]$, which are transported into GRUs network. And the chronological information is extracted and reinforced via this module. The calculation process is formulated as followed:
\begin{equation}
\begin{split}
\centering
  &\textbf{z}_t^u = \sigma(\textbf{W}_u[\textbf{hid}_{t-1},\textbf{h}'_t]+\textbf{b}_u)\\
  &\textbf{z}_t^r = \sigma(\textbf{W}_r[\textbf{hid}_{t-1},\textbf{h}'_t]+\textbf{b}_r)\\
  &\tilde{\textbf{hid}}_t = tanh(\textbf{W}_{\tilde{hid}_t}[\textbf{z}_t^u\otimes\textbf{hid}_{t-1},\textbf{h}'_t]+\textbf{b})\\
  &\textbf{hid}_{t} = \textbf{z}_t^u\otimes\textbf{hid}_{t-1}+(1-\textbf{z}_t^u)\otimes\tilde{\textbf{hid}}_t
\end{split}
\end{equation}
where $\textbf{z}_t^u$ is the update gate unit, $\textbf{z}_t^r$ is the reset gate unit, $\sigma$ is the sigmoid function, $\otimes$ is the element-wise product, $\textbf{W}_u$, $\textbf{W}_r$ and $\textbf{W}_{\tilde{hid}_t}$ are parameters matrices of GRU network. The representation $\textbf{o}_1$ of the series of user histories is the final hidden state of this network, i.e., $\textbf{o}_1 = \textbf{hid}_{k}$.
\subsection{User Preference Learning}
In terms of the extraction of user representation, we use an existing neural network approach, i.e., the multi-head self-attention mechanism, to capture correlations among user's history browsing. The representation of the $i_{th}$ history learned by the $l_{th}$ attention head. Donate the representation of user $\textbf{o}_1$ as $\textbf{u}$. This is formulated as follows:
\begin{equation}
\begin{split}
    &\alpha_{i,j}^l={softmax}({\textbf{u}_i^T\textbf{Q}\textbf{u}_j})\\
    &\textbf{h}_{i,k}^n = \textbf{V}(\sum_{j=1}^{m}\alpha_{i,j}^l\textbf{u}_j)\\
\end{split}
\end{equation}
where $\textbf{Q}$ and $\textbf{V}$ are parameters matrices of self-attention network, and $\alpha_{i,j}^l$ is the representation of the interaction between $i_{th}$ and $j_{th}$ user's history. The multi-head is the concatenation of the output by $h$ self-attention heads, i.e., $\textbf{h}_i^n = [\textbf{h}_{i,1}^n;\textbf{h}_{i,2}^n;...;\textbf{h}_{i,h}^n]$. Likewise, additive attention weights is computed as follows:
\begin{equation}
\begin{split} 
    & w_i^n = \textbf{q}^T_ntanh(\textbf{V}' \times \textbf{h}_i^n + \textbf{b}_n)\\
    &\beta_i^n = softmax(w_i^n)
\end{split}
\end{equation}
where $\textbf{q}$, $\textbf{V}'$, $\textbf{b}_n$ are parameters matrices of self-attention network, and the final representation of user is the weighted summation of the overall histories, which is formulated as follows:
\begin{equation}
    \textbf{o}_2 = \sum_{i=1}^{n}\beta_i^n\textbf{h}_i^n
\end{equation}
\subsection{Model Training}
For online news recommendation platforms where user and news representations can be calculated beforehand , the evaluation should be as simple and efficient as possible to minimise latency \cite{wu2019neural2}. Inspired by the work of Okura et al. \cite{okura2017embedding}, we use a weighted sum to calculate the possibility of a news click. Denote the representation of current behaviours as $\textbf{o}_1$ and previous user behaviours as $\textbf{o}_2$ and the representation of a candidate news $n_z$ as $\textbf{e}_z$, the probability score as $score_z = (\alpha\textbf{o}_1)^T\textbf{e}_z+[(1-\alpha)\textbf{o}_1]^T\textbf{e}_z$. Motivated by Wu et al. \cite{wu2019neural}, we use the random negative sampling method and cross entropy loss during training, which can be formulated as:
\begin{equation}
-\sum_{i=1}^{P}log\frac{exp(score_{z,i}^+)}{exp(score_{z,i}^+)+\sum_{i=1}^{K}exp(score_{z,i}^-)}
\end{equation}
where $P$ is the number of positive samples, and $K$ is the number of negative samples.
\section{Dataset}
\begin{figure}[htb]
\centering
\subfigure[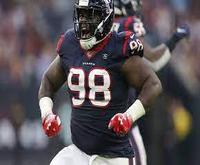]{
\centering
\includegraphics[width=0.4\linewidth]{N1.jpg}
\label{fig:2.a}
}\subfigure[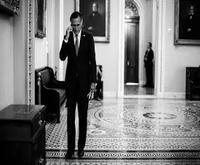]{
\centering   
\includegraphics[width=0.4\linewidth]{N2.jpg}
\label{fig:2.b}
}\caption{Illustration of two examples and their titles. (a): N1: "Texans defensive tackle D.J. Reader is taking advantage of his opportunities." (b) N2: "Mormons to the Rescue?"}
\label{fig:2}  
\end{figure}
\subsection{Dataset Establishment}
To the best of our knowledge, the previous work did not have suitable dataset to support multimodal news recommendation tasks with images and text. Xun et al. \cite{xun2021we} made a great contribution to MIND \cite{wu2020mind} and constructed the IM-MIND \cite{xun2021we} dataset with 41.74\% image coverage. However, for multimodal news recommendation a large number of blank images will affect the robustness of the training process. Since that, we built a benchmark dataset\footnote{The dataset will be released.} on the basis of the MIND dataset \cite{wu2020mind}. Specifically, we used the text information and added the corresponding image information. Examples are illustrated in appendix.

Owing to part of invalid news URLs, we need a different approach to crawl the image of news. It occurred to us that we could search the news with invalid URLs for cover pictures using headlines of the required news in bulk with the HTML
\begin{table*}[ht]
\centering
\begin{tabular}{lccccc}
\hline
    Dataset &  News & Users & Image  & $Image_{rate}$ & $origin_{rate}$ \\
\hline
    MIND-large & 130,379 & 876956 & 0 & 0\% & 0\%\\
    IM-MIND & 130,379 & 876956 & 54421 & 41.74\% & 41.74\%\\
    \textbf {V-MIND} & 130,379 & 876956 & \textbf{130379} & \textbf{100\%} & \textbf{98.92\%}\\
\hline
\end{tabular}
\caption{\label{tab:det}Detailed statistics of MIND, IM-MIND and V-MIND}
\end{table*}
\begin{table*}
\centering
\begin{tabular}{lcccc}
\hline
    Model & AUC & NDCG@5 & NDCG@10 & MRR\\
\hline
    DeepFM & 0.659  & 0.345 & 0.407 & 0.314\\
    DKN & 0.672 & 0.353 & 0.417 & 0.321\\
    NPA & 0.675 & 0.358 & 0.422 & 0.326\\
    LSTUR & 0.680 & 0.363 & 0.427 & 0.323\\
    NRMS & 0.676 & 0.356 & 0.422 & 0.323\\
    FIM & 0.685 & 0.368 & 0.431 & 0.331\\
    NRMS-IM & 0.687 & 0.369 & 0.432 & 0.331\\
    FIM-IM & 0.691 & 0.373 & 0.436 & 0.336\\
\hline
    \textbf{VLSNR(Ours)} & \textbf{0.695}& \textbf{0.376} & \textbf{0.440} & \textbf{0.340}\\
\hline
\end{tabular}
\caption{\label{tab:freq}
Performance Comparison with state-of-the-art news recommendations under V-MIND and MIND-large.
}
\end{table*}
code in Google Images\footnote{https://images.google.com/}. Crawling news images are prioritized from the original site. For the news that is no longer available, we searched for the same news story on other sites in order of similarity of headlines and get its cover image.  After that, we named each figure with the corresponding news id, as shown in figure \ref{fig:2}. To better support future research, all of the images were resized into 224*224px respectively. We have constructed the large dataset, according to MIND-large version.

\subsection{Detailed Statistics}
Table \ref{tab:det} lists the comparison details of MIND, IM-MIND and our V-MIND. The entire dataset contains 130,379 news items and 876,956 users. \textbf{Image} is used to denote the number of news that contains the image. V-MIND contains 130,379 images, which ensures that every news item has a set of text-image information. 1401 news could not get the cover of the original news, so we use the same one in other reports to replace it. The vision coverage of our V-MIND reached 100\%, and the original cover restoration rate reaches 98.92\%.

\section{Experiments}
\subsection{Experimental Settings}
In our experiment, we used the pretrain CLIP encoder, \cite{radford2021learning} as the initialisation of word embedding, i.e., BERT model \cite{devlin2018bert} and image embedding, i.e., Vision Transformers. Both word and image embedding dimension are 512. The head of crossmodal attention network is 8. We applied dropout \cite{srivastava2014dropout} to mitigate overfitting. The dropout rate is 0.5. We used Adam \cite{kingma2014adam} to optimize our model, and the learning rate is 1e-4. The batch size is 256. The number of negative samples for each positive one is 3. We repeated each experiment 5 times, and used universal ranking metrics to evaluate the performance, including AUC, MRR, NDCG@5 and NDCG@10 scores.
\subsection{Performance Comparison}
In this section, we compare the proposed VLSNR model with several baseline methods.We document the performance of different methods on the V-MIND dataset, including: (1) DeepFM \cite{guo2017deepfm}, a popular neural recommendation method which synthesizes deep neural networks and factorization machines; (2) DKN \cite{wang2018dkn}, using CNN to embedding news with both word and entity; (3) NPA \cite{wu2019npa}, learning news representation with personalized attention mechanism;(4) LSTUR \cite{an2019neural}, learning click trends with long-term and short-term user interests; (5) NRMS \cite{wu2019neural2}, using multi-head attention to learn users behavior in news option; (6)  FIM \cite{wang2020fine}, learning news features with a fine-grained method; (7) NRMS-IM and FIM-IM \cite{xun2021we}, using resnet101 ot obtain the feature map of images. In method (1-6), news tests information are the only feature be learned. Method (7-8) add a block for news cover extraction after text. Table \ref{tab:freq} shows the results of the comparison experiment. It shows that methods that combine visual and textual information outperform other methods that consider only text content. This is because users not only focus their interest on the text before clicking on the news, but also depends on how much they like the images. Therefore, we believe that multimodal modeling of text and images in news recommendation systems is more consistent with the simulation of user behavior, which obviously improves the performance of the recommendation system. Our VLSNR achieves better performance in multimodal news recommendations. 

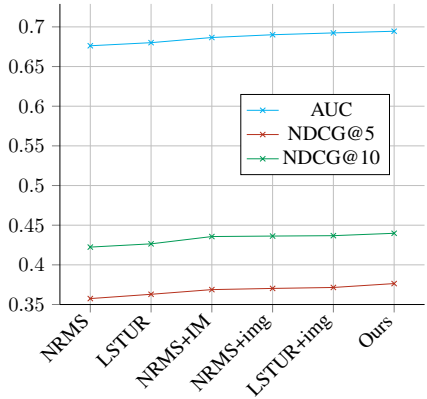
\begin{figure}[!t]
\centering
    \begin{minipage}[t]{0.8\linewidth}
    \begin{tikzpicture}[scale=0.7]
          % \centering
            \begin{axis}[
                ymin=0.35,
                symbolic x coords={NRMS,LSTUR,NRMS+IM,NRMS+img,LSTUR+img,Ours}, xtick={NRMS,LSTUR,NRMS+IM,NRMS+img,LSTUR+img,Ours},
               x tick label style={rotate=45, anchor=east, align=center}, ytick={0.35,0.40,0.45,0.50,0.55,0.60,0.65,0.70},
                axis lines*=left,
                ymajorgrids = true,
                xmajorgrids = true,
                tick align=outside,
                legend style={at={(0.7,0.7)},anchor=north,legend columns=1},
                ]   
                \addplot[draw=ProcessBlue,mark=x] 
                coordinates {
                    (NRMS,0.6762)
                    (LSTUR,0.6801)
                    (NRMS+IM,0.6866)
                    (NRMS+img,0.6901)
                    (LSTUR+img,0.6924)
                    (Ours,0.6945)
                };
                \addlegendentry{AUC}
                \addplot[draw=BrickRed,mark=x] 
                coordinates {
                    (NRMS,0.3575)
                    (LSTUR,0.3629)
                    (NRMS+IM,0.3688)
                    (NRMS+img,0.3703)
                    (LSTUR+img,0.3715)
                    (Ours,0.3764)
                };
                \addlegendentry{NDCG@5}
                \addplot[draw=ForestGreen,mark=x] 
                coordinates {
                    (NRMS,0.4224)
                    (LSTUR,0.4265)
                    (NRMS+IM,0.4357)
                    (NRMS+img,0.4363)
                    (LSTUR+img,0.4368)
                    (Ours,0.4398)
                };
            \addlegendentry{NDCG@10}
            \end{axis}
\end{tikzpicture}
\end{minipage}%
\caption{Effectiveness of different methods}
\label{fig:4}
\end{figure}
Then we note that the temporal information of the user's interest can play a significant role in the prediction of next clicks. We compared A with other methods to show the performance of time-aware. The results are shown in Figure \ref{fig:4}. LSTUR \cite{an2019neural} uses GRU to learn time-aware in a text-only news recommendation model, which shows better performance compared to NRMS. Based on Xun's \cite{xun2021we} contribution of adding image information to NRMS, we replaced the image data with our V-MIND and concatted the visual information on the output vector of LSTUR. We find that the temporal features can learn the fusion information of image and text very well, and VLSNR can show better performance in the fusion of multimodal and temporal features.

\subsection{User Modelling}
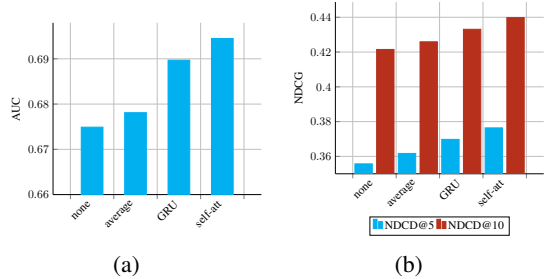
\begin{figure}[htbp]
    \centering
    \subfigure[]{
    	\label{fig:}
        \begin{minipage}[t]{0.47\linewidth}
            \centering
            \begin{tikzpicture}[scale=0.4]
                \centering
                \begin{axis}[
            	    symbolic x coords={none, average, GRU, self-att, 0},
            	    xtick={none, average, GRU, self-att},
            	    x tick label style={rotate=45, anchor=east, align=center},
            	    ylabel=AUC,
            	    axis lines*=left,
            	    ymajorgrids = true,
            	    ybar interval = 0.5,
            	    ymin = 0.66,
            	]
                	\addplot[draw=ProcessBlue, fill=ProcessBlue] 
                	coordinates{
                		(none,0.6749)
                        (average,0.6781)
                        (GRU,0.6897)
                        (self-att,0.6945)
                        (0,0.66)
                	};
                \end{axis}
            \end{tikzpicture}
        \end{minipage}%
    }%
    \subfigure[]{
    	\label{fig:B}
        \begin{minipage}[t]{0.47\linewidth}
            \centering
            \begin{tikzpicture}[scale=0.4]
                \centering
               \begin{axis}[
            	    symbolic x coords={none, average, GRU, self-att, 0},
            	    xtick={none, average, GRU, self-att},
            	    x tick label style={rotate=45, anchor=east, align=center},
            	    ylabel=NDCG,
            	    axis lines*=left,
            	    ymajorgrids = true,
            	    ybar interval = 0.8,
            	    ymin = 0.35,
            	    legend style={at={(0.5,-0.25)},anchor=north,legend columns=-1},
            	]
                	\addplot[draw=ProcessBlue, fill=ProcessBlue] 
                	coordinates{
                		(none,0.3557)
                        (average,0.3617)
                        (GRU,0.3698)
                        (self-att,0.3764)
                        (0,0.35)
                	};
                	\addplot[draw=BrickRed, fill=BrickRed] 
                	coordinates{
                		(none,0.4215)
                        (average,0.4259)
                        (GRU,0.4331)
                        (self-att,0.4398)
                        (0,0.35)
                	};
                	\addlegendentry{NDCD@5}
                    \addlegendentry{NDCD@10}
                \end{axis}
            \end{tikzpicture}
        \end{minipage}%
    }%
\caption{Ablation studies on different user modelling.} 
\label{fig:7}
\end{figure}

User modelling is core of the design of personalised news recommendation systems. Therefore, it is essential to investigate the performance differences among various modelling approaches. We present several ablation studies on different methods, including none user embedding, weighted average of the representation of clicked histories, GRU network \cite{okura2017embedding}, self-attention mechanism \cite{wu2019neural2}. The performances are shown in Figure \ref{fig:7}.
Apparently, user encoders can improve performance, compared to none user encoder. Then, GRU and self-attention both outperform average method. This may be attributed to the fact that the averaging method does not make use of information on timing or user attention, which can be informative when we model users based on their historical browsing. When it comes to self-attention approach, the best performance among the four, its global focus on long sequences makes it superior to GRU in user modelling.
\subsection{Vision-Linguistics Ablation Study}
\begin{figure}[!t]
\centering
    \begin{minipage}[t]{0.8\linewidth}
    \begin{tikzpicture}[scale=0.7]
          % \centering
            \begin{axis}[
                ylabel= AUC,
                xlabel = proportion,
                ymin=0.67, xtick={0,0.1,0.2,0.3,0.4,0.5,0.6,0.7,0.8,0.9,1}, ytick={0.67,0.675,0.68,0.685,0.69,0.695,0.70},
                axis lines*=left,
                ymajorgrids = true,
                xmajorgrids = true,
                tick align=outside,
                legend style={at={(0.7,0.3)},anchor=north,legend columns=1},
                ]   
                \addplot[draw=ProcessBlue,mark=x] 
                coordinates {
                    (0,0.6781)
                    (0.1,0.6795)
                    (0.2,0.6813)
                    (0.3,0.6833)
                    (0.4,0.6865)
                    (0.5,0.6898)
                    (0.6,0.6928)
                    (0.7,0.6933)
                    (0.8,0.6949)
                    (0.9,0.6937)
                    (1,0.6945)
                };
                \addlegendentry{Ours}
                \addplot[draw=BrickRed,mark=x] 
                coordinates {
                    (0,0.6741)
                    (0.1,0.6757)
                    (0.2,0.6764)
                    (0.3,0.6796)
                    (0.4,0.6821)
                    (0.5,0.6848)
                    (0.6,0.6888)
                    (0.7,0.6889)
                    (0.8,0.6899)
                    (0.9,0.6902)
                    (1,0.6901)
                };
                \addlegendentry{NRMS+img}
                \addplot[draw=ForestGreen,mark=x] 
                coordinates {
                    (0,0.6765)
                    (0.1,0.6782)
                    (0.2,0.679)
                    (0.3,0.6813)
                    (0.4,0.6845)
                    (0.5,0.6888)
                    (0.6,0.69)
                    (0.7,0.6913)
                    (0.8,0.6929)
                    (0.9,0.6917)
                    (1,0.6925)
                };
            \addlegendentry{LSTUR+img}
            \end{axis}
\end{tikzpicture}
\end{minipage}%
\caption{Ablation studies on different image proportion} 
\label{fig:5}
\end{figure}
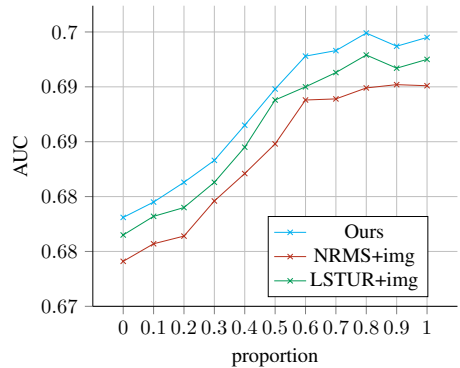

Then, we study the effectiveness of multimodal information on news recommendation through ablation experiments.  We adjust the ratio of the image information in the dataset during the training process. Another part of the images is replaced with 224*224px white images, and the text information is the same. The results are shown in Figure \ref{fig:5}. We find that the performance of the recommendation system grows as the proportion of news with visual features increases. AUC rises slowly after the image ratio reaches 70\% and eventually reaches its peak, which shows that visual information can improve the performance of news recommendation. 

\begin{table}
\centering
\resizebox{\linewidth}{!}{
\begin{tabular}{lccc}
\hline
    AUC & NRMS+img & LSTUR+img & \textbf{Ours}\\
\hline
    top & 0.621 & 0.617 & \textbf{0.619}\\
    sub & 0.637 & 0.633 & \textbf{0.635}\\
    tit & 0.667 & 0.663 & \textbf{0.664}\\
    tit+top+sub & 0.672 & 0.667 & \textbf{0.670}\\
    tit+img & 0.681 & 0.677 & \textbf{0.680}\\
    overall & 0.694 & 0.690 & \textbf{0.692}\\
\hline
\end{tabular}
}
\caption{\label{tab:3}Ablation studies on different news information}
\end{table}
In order to better study the influence of news composition on recommendation results, we split the text information of news and images for a comprehensive comparison. Table \ref{tab:3} shows that semantic contribution of title in text information is greater than that topic and subtopic. News recommendations perform better when image information is combined with title. The best performance is achieved when the visual information of the news and all text information are multimodally fused in VLSNR.
\begin{figure}[htbp]
    \centering
    \subfigure[]{
    	\label{fig:AA}
        \begin{minipage}[t]{0.47\linewidth}
            \centering
            \begin{tikzpicture}[scale=0.4]
                \centering
                \begin{axis}[
            	    symbolic x coords={basic, fine-tune,0},
            	    xtick={basic, fine-tune},
            	    x tick label style={rotate=45, anchor=east, align=center},
            	    ylabel=AUC,
            	    axis lines*=left,
            	    ymajorgrids = true,
            	    ybar interval = 0.5,
            	    ymin = 0.67,
            	]
                \addplot[draw=ProcessBlue, fill=ProcessBlue] 
                coordinates{
                	(basic,0.6721)
                    (fine-tune,0.6945)
                    (0,0.67)
                };
                \end{axis}
            \end{tikzpicture}
        \end{minipage}%
    }%
    \subfigure[]{
    	\label{fig:BB}
        \begin{minipage}[t]{0.47\linewidth}
            \centering
            \begin{tikzpicture}[scale=0.4]
                \centering
               \begin{axis}[
            	    symbolic x coords={basic, fine-tune,0},
            	    xtick={basic, fine-tune},
            	    x tick label style={rotate=45, anchor=east, align=center},
            	    ylabel=NDCG,
            	    axis lines*=left,
            	    ymajorgrids = true,
            	    ybar interval = 0.8,
            	    ymin = 0.35,
            	    legend style={at={(0.5,-0.25)},anchor=north,legend columns=-1},
            	]
                	\addplot[draw=ProcessBlue, fill=ProcessBlue] 
                	coordinates{
                		(basic,0.3539)
                        (fine-tune,0.3764)
                        (0,0.35)
                	};
                	\addplot[draw=BrickRed, fill=BrickRed] 
                	coordinates{
                		(basic,0.4198)
                        (fine-tune,0.4398)
                        (0,0.35)
                	};
                	\addlegendentry{NDCD@5}
                    \addlegendentry{NDCD@10}
                \end{axis}
            \end{tikzpicture}
        \end{minipage}%
    }%
\caption{Ablation studies on different user modelling.} 
\label{fig:8}
\end{figure}
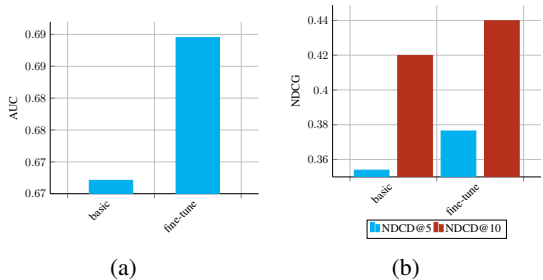

We find that there are some performance limitations in clip’s pre-trained model, such as poor results on finer-grained recognition
and low accuracy in understanding abstract news headlines. We therefore further trained clip’s model and installed it into our model to participate in the training together, which eventually achieved better results, as illustration in Figure \ref{fig:8}. Fine-tuning of the CLIP model has resulted in good performance improvements.
\subsection{Ranking Contrast}
\begin{figure}[htb]
\centering
\includegraphics[width=\linewidth]{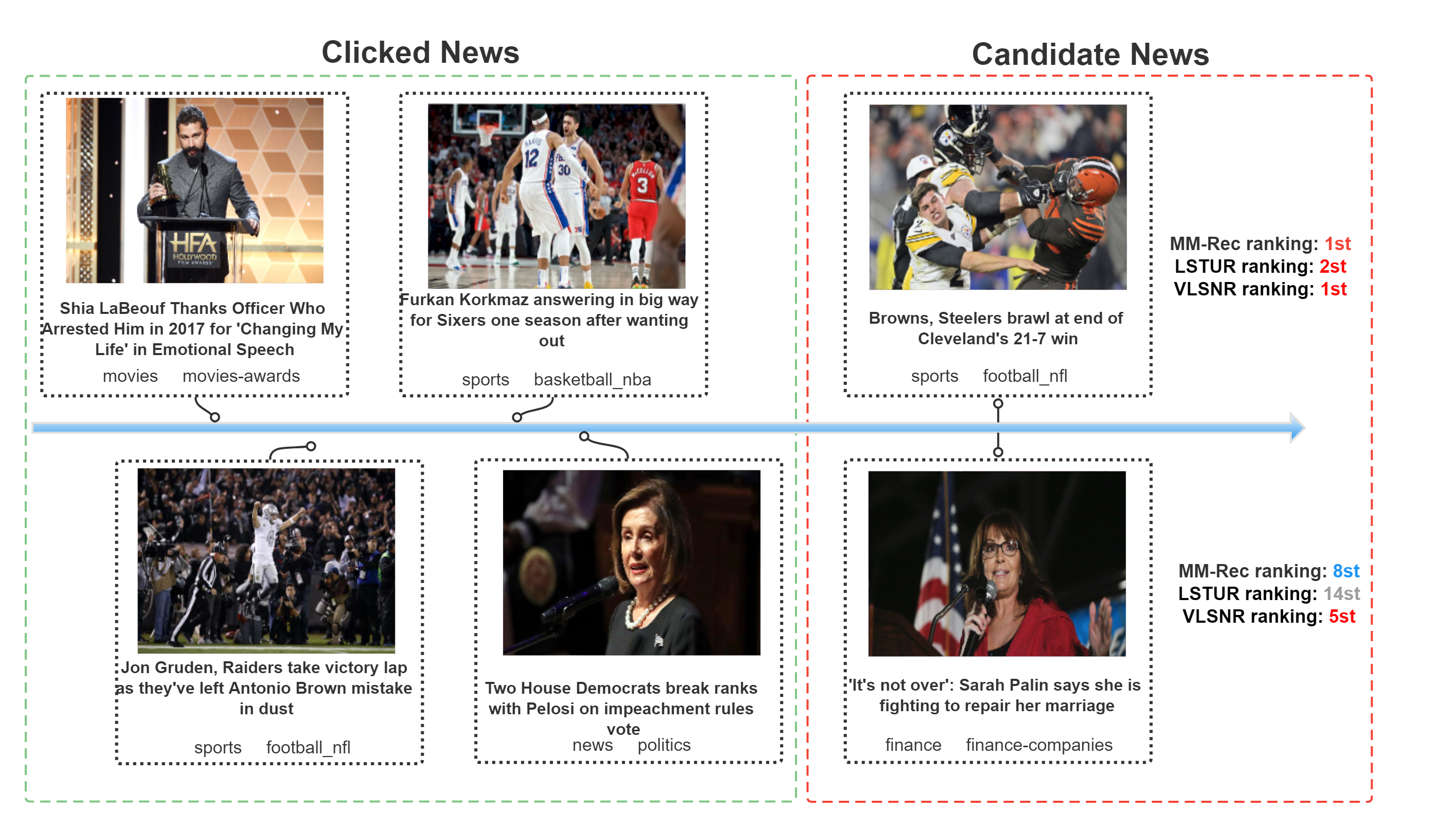}
\caption{An example of a user's browsing history and candidate news. The predicted ranking of MM-Rec, LSTUR and VLSNR(Ours) is also shown on the graph.} 
\label{fig:10}
\end{figure}

Motivated by MM-Rec\cite{wu2021mm}, we are curious about the different methods (MM-Rec, LSTUR\cite{wu2019neural} and \textbf{VLSNR}) that generated predictions of user click behaviour. We therefore study a set of historical reading behaviours of a user. Results are displayed in Figure \ref{fig:10}. We find that this user viewed news on sports (including rugby and basketball), movies and politics. One interesting phenomenon is that the predictions for the first piece of candidate news are strikingly similar. This is probably because historical information such as "Raiders", "Sixers" and the corresponding multimodal information indicates that the user is a sports fan. However, for another candidate, the results are quite diverse. LSTUR is completely incapable to comprehend the relevance of this news to previous news, which is to be expected, as it cannot make use of visual information. Both multimodal approaches capture the association between the two women in politics. It is worth noting that our method sorts this news further up the list due to time sensitivity.
These results show the effectiveness of crossmodal information and the significance of time-sequence in news recommendation.
\section{Conclusion and Future Work}
In this paper we propose VLSNR, which can use fused information from images and text for news recommendation. We encode text and images through tune's clip training model and capture important features between images and text through a layer of crossmodal attention and a layer of additive attention. We have improved the time series model so that it learns better about user features, which is an important addition to previous work. We contribute the MIND dataset with visual information to promote more researches on news recommendation in a multimodal fields. Extensive experiments have shown that VLSNR can effectively improve the performance of news recommendation because of its extraction of multimodal information and perception of time series. 

In subsequent work we will explore and use better multimodal models, such as those with greater text analysis capability and finer granularity, and seek to pre-train a large number of stories to achieve better results.

\section*{Limitations}
Firstly, the multimodal model we use is an existing model that works well, however, the problems with the model (including those related to granularity) limit our work to some extent. Then, we work only in the English semantic context. As multimodal datasets for news recommendations are relatively scarce, we can only test from our dataset as well.

\bibliography{anthology,custom}
\bibliographystyle{acl_natbib}

\newpage
\appendix
\end{document}